\documentclass[journal]{IEEEtran}
\usepackage{array,multirow}
\usepackage{graphicx}
\usepackage{tabularx}
\usepackage{tabulary}
\usepackage{latexsym}
\usepackage{textgreek}
\usepackage{layout}
\usepackage{rotating}
\usepackage{amsmath,amsfonts,amssymb}
\usepackage{bm}
\usepackage{euscript}
\usepackage{algorithmic}
\usepackage{amssymb, amsmath}
\usepackage[numbers, square, comma, sort&compress]{natbib}
\usepackage[ruled,lined]{algorithm2e}
\usepackage{epstopdf}
\usepackage{gensymb}
\usepackage{subcaption}
\usepackage{siunitx}
\usepackage{url}
\usepackage{xcolor}
\usepackage{ragged2e}
\usepackage[english]{babel}
\usepackage{multicol}
\usepackage[official]{eurosym}
\usepackage{color}
\usepackage{acronym}
\usepackage{mathtools}
\usepackage[verbose]{wrapfig}
\usepackage{multirow}
\usepackage[normalem]{ulem}

\usepackage [autostyle, english = american]{csquotes}
\MakeOuterQuote{"}
\usepackage{caption} 
   \usepackage[belowskip=2pt, aboveskip=-13pt] {caption}
\DeclarePairedDelimiter{\ceil}{\lceil}{\rceil}

\newcommand{\eq}[1]{(\ref{#1})}

\acrodef{API}[API]{Application Programmable Interfaces}
\acrodef{BS}[BS]{Base Station}
\acrodef{CNN}[CNN]{Convolutional Neural Network}
\acrodef{CPU}[CPU]{Central Processing Unit}

\acrodef{EB}[EB]{Energy Buffer}
\acrodef{EH}[EH]{Energy Harvesting}
\acrodef{ES}[ES]{Energy Saving}
\acrodef{EM}[EM]{Energy Manager}
\acrodef{GENM}[GENM]{Green-based Edge Network Management}
\acrodef{EPC}[EPC]{Evolved Packet Core}
\acrodef{ETSI}[ETSI]{European Telecommunications Standards Institute}
\acrodef{GP}[GP]{Geometric Programming}
\acrodef{ITS}[ITS] {Intelligent Transport System}
\acrodef{LOC}[LOC]{User Location Services} 
\acrodef{LLC}[LLC]{Limited Lookahead Control}
\acrodef{LS}[LS]{Location Service}
\acrodef{LSTM}[LSTM]{Long Short-Term Memory}
\acrodef{MEC}[MEC]{Multi-access Edge Computing}
\acrodef{ML}[ML]{Machine Learning}
\acrodef{MN}[MN]{Mobile Network}
\acrodef{TIM}[TIM]{Telecom Italia Mobile}

\acrodef{NOES}[NOES]{NO Energy Saving}
\acrodef{NFV}[NFV]{Network Function Virtualization}
\acrodef{NIC}[NIC]{Network Interface Card}
\acrodef{QoS}[QoS]{Quality of Service}
\acrodef{RNN}[RNN]{Recurrent Neural Network}
\acrodef{RAN}[RAN]{Radio Access Network}
\acrodef{RMSE}[RMSE]{Root Mean Square Error}
\acrodef{RNN}[RNN]{Recurrent Neural Network}
\acrodef{UE}[UE]{User Equipment}
\acrodef{VM}[VM] {Virtual Machine}
\acrodef{VNF}[VNF]{Virtualized Network Function}

\begin{document}

\title{LSTM-based Traffic Load Balancing and Resource Allocation for an Edge System}



\author{\IEEEauthorblockN {Thembelihle Dlamini\IEEEauthorrefmark{1}, Sifiso Vilakati \IEEEauthorrefmark{2}}\\
	\IEEEauthorblockA {\IEEEauthorrefmark{1}Department of Electrical and Electronic Engineering, University of eSwatini, Kwaluseni, Eswatini}\\
	\IEEEauthorblockA {\IEEEauthorrefmark{2}Department of Statistics and Demography, University of eSwatini, Kwaluseni, Eswatini}\\
	\{tldlamini, svilakati\}@uniswa.sz
	 \vspace{-0.4cm}
}

\maketitle
\thispagestyle{plain}
\pagestyle{plain}


\begin{abstract}

The massive deployment of small cell Base Stations (SBSs) empowered with computing capabilities presents one of the most ingenious solutions adopted for $5$G cellular networks towards meeting the foreseen data explosion and the \mbox{ultra-low} latency demanded by mobile applications. This empowerment of SBSs with \textit{\mbox{Multi-access} Edge Computing (MEC)} has emerged as a tentative solution to overcome the latency demands and bandwidth consumption required by mobile applications at the network edge. The MEC paradigm offers a limited amount of resources to support computation, thus mandating the use of intelligence mechanisms for resource allocation. The use of green energy for powering the network apparatuses (e.g., Base Stations (BSs), MEC servers) has attracted attention towards minimizing the carbon footprint and network operational costs. However, due to their high intermittency and unpredictability, the adoption of learning methods is a requisite. Towards intelligent edge system management, this paper proposes a \mbox{Green-based} Edge Network Management (GENM) algorithm, which is a online edge system management algorithm for enabling \mbox{green-based} load balancing in BSs and energy savings within the MEC server. The main goal is to minimize the overall energy consumption and guarantee the Quality of Service (QoS) within the network. To achieve this, the GENM algorithm performs dynamic management of BSs, autoscaling and reconfiguration of the computing resources, and on/off switching of the fast tunable laser drivers coupled with \mbox{location-aware} traffic scheduling in the MEC server. The obtained simulation results validate our analysis and demonstrate the superior performance of GENM compared to a benchmark algorithm.
 
\end{abstract}

\begin{IEEEkeywords}
	Multi-access edge computing, green energy, autoscaling, sustainability, machine learning. 
\end{IEEEkeywords}

\IEEEpeerreviewmaketitle

\section{Introduction}

The foreseen {\it dense} deployment of \acp{BS} empowered with computing capabilities in order to meet the \mbox{ultra-low} latency demanded by mobile users raises concerns related to energy consumption. Apart from the fact that \ac{BS} energy costs accounts for a large part of the operating expenses of \ac{MN} operators, there are also increasing concerns regarding their environmental impact in terms of high carbon dioxide ($\rm CO_2$) emissions. In an effort to minimize energy consumption and energy costs in $5$G cellular networks within the \ac{MEC} paradigm, this paper advocates for the integration of \ac{EH} systems into network apparatuses and the use of \mbox{container-based} virtualization within computing platforms (i.e., MEC servers). 
The use of green energy mitigates the negative environmental impact of \acp{MN} and enable cost saving for mobile operators in terms of lowering  operational energy costs. The motivation towards green energy is due to the fact that current trends in battery and solar module costs show a reduction in prices.
The benefits of \mbox{container-based} virtualization is the reduction in energy drained in the computing platform due to their lower overheads when compared with \acp{VM}~\cite{morabito_container}\cite{interdigital}. For a qualitative comparison of different virtualization technniques, interested readers are referred to~\cite{morabito_container}.

In this paper, a group of \ac{EH} \acp{BS} placed in proximity to a \mbox{EH-MEC} server are considered as an {\it edge system}. The \mbox{EH-MEC} server manages the BS system, deciding upon the allocation of transmission resources, and also handling the computing and communication processes.
In general, renewable energy systems are dimensioned to guarantee the autonomous operation of the edge system. Thus, it is desirable that the utilization of green energy be made one of the performance metrics when designing load-balancing strategies~\cite{green_balancing}\cite{globe_balancing}, instead of the network impact~\cite{E_oh}\cite{edge_controller}. As a result, the \mbox{green-based} load balancing is a promising technique for optimizing \ac{MEC} performance since it exploits the spatial diversity of the available green energy to reshape the  network load among the BSs~\cite{globe_balancing}. In this case, \acp{MN} can prioritize the utilization of \acp{BS} with sufficient green energy to serve more traffic while reducing the traffic loads for those \acp{BS} consuming \mbox{on-grid} power. For instance, in the \ac{MEC} server, a \mbox{trade-off} between the green energy utilization and the amount of workload that can be computed locally should be carefully evaluated. 

In this regard, it is worth noting that the energy consumption within the virtualized computing platform is due to (i) the active computing resources, i.e., \acp{VM} or containers~\cite{virttech}\cite{eempirical}\cite{tbook}, (ii) the network communications, communication related to transmission drivers~\cite{link_drivers}\cite{comp_plus_comm_mec}, and the \mbox{intra-communications}~\cite{vm_book}. In order to alleviate this, this paper assumes that the \mbox{container-based} virtualization be deployed in the MEC server as containers are lightweight, i.e., demand less memory space, have shorter \mbox{start-up} time, and offer software portability. At each time instance, the containers are provisioned based on the forecasted server workloads, a technique referred to as \mbox{\it autoscaling}. In addition, the transmission drivers used for data transfers within the server are tuned by taking into account mobile user's location. The server is also capable of caching the frequently requested contents locally.

In \mbox{densely-deployed} \acp{BS}, the energy drained is due to the \mbox{always-on} design approach~\cite{E_oh}\cite{edge_controller}\cite{oh2011toward}, yet traffic load varies during the day, e.g., the demand is low during the night. Therefore, in order to intelligently manage the \acp{BS} towards minimizing the energy consumption, the \mbox{green-based} load balancing technique is employed, i.e., BS sleep modes are enabled in some \acp{BS} using green energy as a performance metric.

\textbf{Paper contributions:} this paper considers an energy cost model that takes into account the computing, caching and communication processes within the \ac{MEC} server, and  \mbox{transmission-related} energy consumption in \acp{BS}. Here, the GENM algorithm is proposed for enabling \mbox{\it green-based} traffic load balancing, i.e., the \acp{BS} are dynamically switched on/off based on their harvested energy, autoscaling and reconfiguring the computing resources, and the tuning of transmission drivers. This entails using a minimum number of optical drivers for \mbox{real-time} data transfers, over a \mbox{short-term} horizon. In order to solve the energy consumption minimization problem, the GENM algorithm performs online supervisory control, utilizing the learned traffic load and the harvested energy patterns. Then, the output is utilized within a \ac{LLC} policy~\cite{llcprediction} to obtain the optimal system control actions that yields the desired energy savings that guarantees the required \ac{QoS}.  

This work is an extension of~\cite{comp_plus_comm_mec}, where energy savings and \ac{QoS} guarantee were considered only within a virtualized computing platform placed in proximity to a cluster of \acp{BS}. In~\cite{comp_plus_comm_mec}, the role of the \ac{MEC} server is to handle the offloaded computational workload only, which means that the energy cost model lacks the consideration of the BS management procedures, caching process, and the use of containers.

The summary of contributions are listed as follows: 
\begin{itemize}
    \item[1)] The use of \mbox{container-based} virtualization is introduced as they are lightweight, i.e., demand less memory space, have shorter \mbox{start-up} time, and offer software portability.
    \item[2)] The proposed GENM algorithm, which is an online edge management system, makes use of predictive optimization, specifically using the \ac{LLC}, where \mbox{green-based} load balancing, containers provisioning and the tuning of the transmission drivers is performed based on the learned information.  
    \item[3)] The numerical results, obtained with \mbox{real-world} harvested energy and traffic load traces, shows that the proposed optimization strategy is able to efficiently manage the edge network resources in order to minimize the energy drained under the guidance of the intelligent \mbox{online-based} resource manager and the energy saving procedures. 
\end{itemize}

In order to achieve these, the remainder of the paper is organized as follows: Section~\ref{sec:rel_work} describes the related work. Section~\ref{sec:sys} explains the system model. In Section~\ref{sec:prob}, the design and the implementation of the online algorithm is presented. Simulated results are discussed in Section~\ref{sec:results}. Lastly, the work is concluded in Section~\ref{sec:concl}.

\section{Related Work}
\label{sec:rel_work}

\subsection{Methods for load balancing in \acp{MN}}

Load balancing has been studied towards data center management whereby the data center servers employ temporal dependency strategies, i.e., the servers are turned on/off depending on the arrival rates of workloads. This significantly differs from our considered problem as we consider load balancing in SBSs powered by green energy.
Towards load balancing, the dynamic \ac{BS} switching on/off strategies have been used. However, this may have an impact on the network due to the load that is offloaded to the neighboring BS(s). To avoid this, the BS to be switched off is carefully identified within the \ac{BS} cluster. In~\cite{E_oh}\cite{edge_controller}, the {\it network impact} is used to identify the BS to be switched off, one at a time, with no significant network performance degradation. 
Taking into account daily traffic load variation, strategies for opportunistic utilization of the unexploited \mbox{third-party} small cells (SCs) capacity is exploited towards energy savings in~\cite{bousia2016multiobjective}, in order to enable the switching off of some BSs. Here, an offloading mechanism is introduced, where the operators lease the capacity of a SC network owned by a third party in order to switch off their \acp{BS} (Macro \acp{BS}) and maximize their energy efficiency, when the traffic demand is low.

The use of green energy as a performance metric has been explored within the Radio Acess Network (RAN)~\cite{green_balancing}\cite{Han2013}. 
Along the lines of \ac{MN} softwarization, a distributed user association scheme that makes use of the SoftRAN concept for traffic load balancing via the RAN Controller (RANC) is proposed in~\cite{green_balancing}. Here, the user association algorithm runs on the RANC and the users report their downlink data rates via the associated BS to the RANC, where the traffic loads from individual users and \acp{BS} are measured. The algorithm enhance the network performance by reducing the average traffic delivery latency in \acp{BS} as well as to reduce the \mbox{on-grid} power consumption by optimizing the green energy usage.
Then, authors in~\cite{Han2013} proposed to optimize the utilization of green energy for cellular networks by optimizing the BSs transmission power. The proposed scheme achieves significant \mbox{on-grid} power savings by scheduling the green energy consumption along the time domain for individual BS, and balancing the green energy consumption among the BSs. 

Along the lines of MEC, the authors in~\cite{globe_balancing} proposed a framework for jointly performing load balancing, admission control and energy purchase within a network of \mbox{EH-powered} \acp{BS} with the goal of minimizing the computation delay and data traffic drops (i.e., increasing the locally computed workloads). This work  use green energy as a performance metric. To solve this problem, an online and distributed algorithm is proposed leveraging the Lyapunov optimization with perturbation technique. Here, the algorithm makes the traffic load decisions without forecasting the future traffic load and harvested energy. 
In contrast, the work presented in this paper consider the \mbox{short-term} future traffic load and the harvested energy for decision making.
Then, in our previous work~\cite{edge_controller}, a supervisory online control algorithm that make use of clustering and the network impact metric towards load balancing in \ac{MN} is proposed. Here, the BSs are empowered with computation capabilities (with \acp{VM} as computing resources), the \ac{LSTM} neural network is used for forecasting and the \ac{LLC} policy handles foresighted optimization. 
Even though these works perform load balancing, the problem and scenario considered in this paper is different. Here, a MEC server manage the SBSs powered by green energy. Similar to~\cite{comp_plus_comm_mec}, forecasting and foresighted optimization is used for edge system management.

Load balancing that follows the energy routing, i.e., more computational workload is offloaded to where more energy is available, is presented in~\cite{chen2018computation}. To handle spatial uneven computation workloads experienced by the \mbox{MEC-enabled} BSs, the authors proposed a peer offloading scheme. Here, the BSs share their computing resources and energy costs.

\subsection{Methods for energy saving within computing platforms}
 
Green computing over data centers is an emerging paradigm that aims at performing the dynamic energy-saving management of data center infrastructures.
Here, procedures for the dynamic on/off switching of servers have been proposed as a way of minimizing energy consumption in computing platforms. 
A novel \mbox{post-decision} state based learning algorithm for server provisioning at the network edge is presented in~\cite{xu2016online}. This work incorporates green energy. At the beginning of the time slot the servers are consolidated, i.e., the number of turned on physical servers are minimized, using the learned optimal  policy for dynamic workload offloading and the autoscaling (or right-sizing). Then, in our previous works~\cite{edge_controller}\cite{comp_plus_comm_mec}\cite{online_pimrc}, VM \mbox{\it soft-scaling} (i.e., the reduction of computing resources per time instance) is employed towards energy saving in virtualized platforms either energized by only renewable energy or hybrid supplies (solar and power grid). This is achieved by forecasting the traffic load and harvested energy, and then employing foresighted optimization to obtain the system control inputs. The work of~\cite{shojafar2015energy} use an iterative algorithm to obtain the number of computing resources (VMs) to be provisioned within a node that transmit to clients wireless. Then, the work of~\cite{vm_char} consider a vehicular scenario where vehicles connect wireless to Fog nodes and then develop an adaptive scheduler, which computes \mbox{on-the-fly} the solutions of both the resource reconfiguration and consolidation problems. For this purpose, the \mbox{primal-dual} algorithm is used. 

In computing platforms, computation offloading strategies can be jointly exploited together with delay constraints towards energy savings. The authors in~\cite{delay} proposed an offloading policy to find the optimal place where to offload and the amount of offloaded task data. In this work, the time taken for processing the offloaded task is reduced, at the same time consuming less energy. Then in~\cite{jamil2020job}, an efficient scheduling for \mbox{latency-sensitive} applications is proposed towards energy and response time minimization. The achieved results show a reduction in delay and network usage, and the energy consumption. In addition, the works of~\cite{mukherjee2019joint} jointly optimize the computing and communication resources, taking into account the local task execution delay and transmission delay. To meet the task delay requirements, in~\cite{task_coord}, the heterogeneous clouds, i.e., edge and remote cloud, are coordinated. Here, different policies are employed in the clouds. In this, the edge cloud handles tasks with loose delay bounds and drops drops tasks with stringent delay bounds when the traffic load is heavy.

Towards minimizing energy consumption induced by communication activities within a computing node, the idea of tuning transmission drivers, as {\it one} of the energy saving strategies within the \ac{MN} infrastructure, is first conceived in~\cite{link_drivers}\cite{laser_tuning} where a computing node (router) is considered. Here, it is observed that having the least number of data transmission drivers ({\it fast} tunable lasers) can yield significant amount of energy savings. Motivated by the aforementioned works, within the MEC paradigm, the authors in~\cite{comp_plus_comm_mec} put forward a traffic engineering- and MEC \ac{LS}-based algorithm that use a \mbox{location-aware} procedure for provisioning the transmission drivers for data transfer towards target \acp{BS}. Here, the MEC \ac{LS} \ac{API} is employed for retrieving the \ac{UE}'s location and then passing the information to the authorized applications within the \ac{MEC} platform, for decision making.

Overall, these works numerically analyze and test the energy performance of some \mbox{state-of-the-art} schedulers for computing platforms, but do not attempt to optimize it through the dynamic joint scaling of the available \mbox{communication-plus-computing} resources. The joint analysis of the \mbox{computing-plus-communication} energy consumption within the MEC paradigm  is still an open research topic. 

\subsection{Methods for guaranteeing quality of service and enabling energy savings (within the MEC paradigm)}  

The mobile operator's goal is to provide QoS Internet services for large populations of clients, while minimizing the overall \mbox{computing-plus-communication} energy consumption. Hence, a trade-off is required between QoS and energy savings.
Future MNs are expected to learn the diverse characteristics of users behavior, as well as renewable energy variations, in order to autonomously determine good system configurations. Towards this goal, online forecasting using \ac{ML} techniques and the \ac{LLC} method can yield the desired system behavior when taking into account the environmental inputs, i.e., \ac{BS} traffic load, server workloads and energy to be harvested. Next, the mathematical tools that are used in this research work are reviewed, namely the \ac{LLC} method~\cite{llcprediction}\cite{llc_datacenter}\cite{hayes_2004} and \ac{LSTM} neural networks~\cite{forecasting}~\cite{lstmlearn}.

The LLC has been used in~\cite{llc_datacenter} to address a resource provision problem within virtualized environments. The optimization problem is posed as a profit maximization problem under uncertainty and the LLC formulation models the cost of control. To address the uncertainty over the workload arrival, the Kalman filter is used. Then, in~\cite{chung1992limited}, an online supervisory control scheme based on \ac{LLC} policies is proposed. Here, after the occurrence of an event, the next control action is determined by estimating the system behavior a few steps into the future using the currently available information as inputs. The control actions exploration is performed using a search tree assuming that the controller knows all future possible states of the process over the prediction horizon. 
An online control framework for resource management in switching hybrid systems is proposed in~\cite{llcprediction}, where the system's control inputs are finite. The relevant parameters of the operating environment, e.g., workload arrival, are estimated and then used by the system to forecast future behavior over a \mbox{look-ahead} horizon. From this, the controller optimizes the predicted system behavior following the specified \ac{QoS} through the selection of the system controls.
In~\cite{edge_controller} and~\cite{online_pimrc}, a \ac{LLC}-based supervisory algorithm is proposed to obtain the system control actions yielding the desired \mbox{trade-off} between energy consumption and \ac{QoS}. Here, the traffic load and harvested energy is forecasted and then used as input in the algorithm. The BS are \mbox{densely-deployed} in~\cite{edge_controller} and each \ac{BS} is empowered with computation capabilities. Then, in~\cite{online_pimrc}, a remote site powered by only green energy is considered.

\begin{figure} [t]
	\centering
	\includegraphics[width = \columnwidth]{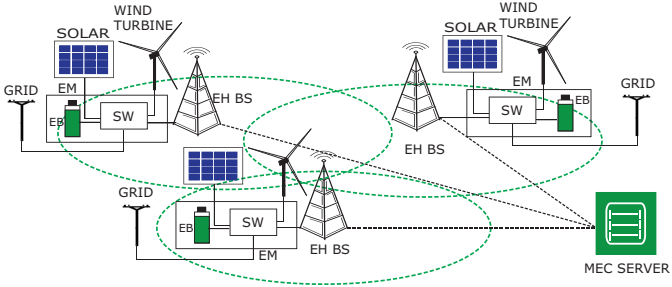}
	\caption{Edge system energized by hybrid energy sources: \mbox{on-grid} power and green energy (solar and wind).}
	\label{fig:eh_mec}
\end{figure} 

\ac{LSTM}  can be used for \mbox{multi-step} time series forecasting as it is able to handle the \mbox{long-term} dependencies due to its inherent capability of storing past information and then recalling it. The forecasting method is utilized in~\cite{edge_controller}\cite{online_pimrc} within an \mbox{LLC-based} algorithm to obtain the system control actions yielding the desired trade-off between energy consumption and QoS. 
The application of \ac{LSTM} network is extended to include \acp{ITS} in~\cite{ferdowsi2017deep}. A new \ac{ITS} edge analytics architecture that makes use of deep learning techniques that either runs on the mobile devices or on the \mbox{intra-vehicle} processors for data analytics is presented. A combination of \ac{LSTM} networks and deep \acp{CNN} is adopted, i.e., \mbox{CNN-LSTM} network, for path selection in autonomous vehicles, whereby the \ac{CNN} is used for feature extraction, and then the extracted information is fed into \ac{LSTM} networks for driving path selection.
Forecasting server workloads using LSTM network can be beneficial for dynamic resource scaling and power consumption in cloud computing datacenters. In~\cite{kumar2018long}, a forecasting model using the LSTM network for predicting future data center workloads is proposed, and then the results are fed into the resource manager for decision making, which either involves scaling up or down the computing resources (servers in this case).

\begin{table}[t]
\footnotesize
\centering
\caption{Notation: list of symbols used in the analysis.}
\begin{tabulary}{1.0\textwidth}{|L|L|}
\hline
{\bf Symbol} & {\bf Description} \\
\hline
\multicolumn{2}{|c|}{\bf Input Parameters} \\
\hline
$C$ & maximum number of containers hosted by the \\ 
    & MEC server, indexed by $c$\\
$N$ & number of \acp{BS}, indexed by $n$\\ 
$\xi(t)$ & aggregate computational workload \\
$\tau$ & time slot duration \\
$L_{\rm in}(t)$ & amount of aggregate workload at the input\\
      & buffer \\
$L_{\rm out}(t)$ & amount of aggregate workload at the output \\
      & buffer \\
$L_{\rm out}^{\rm max}(t), L_{\rm in}^{\rm max}(t)$ & workload buffers maximum capacity\\
$f_{\rm max}$ & maximum processing rate for container $c$\\
$\theta_{{\rm idle}, c}(t)$ & static energy consumed by container $c$ in \\
     & the idle state\\
$\theta_{{\rm max}, c(t)}$  & maximum energy consumed by container $c$ at\\
     & maximum processing rate\\
$z_e$ & \mbox{per-container} reconfiguration cost caused by a \\
     & unit-size frequency switching\\
$\lambda_c(t)$ & workload fraction to be computed by the $c$-th\\
    &  container\\
$\lambda_{\max}$ & maximum computation load \mbox{per-container} \\
$\Delta $ & maximum \mbox{per-slot} and \mbox{per-container} allowed\\
        &  processing time\\
$\theta_{\rm idle}^{\rm NIC}(t)$ & energy drained by the NIC when powered, \\
        & with no data transfer \\
$M$ & maximum number of multiple fast tunable lasers\\
$\beta_{\rm max}$ & maximum energy buffer capacity \\
$\beta_{\rm up}, \beta_{\rm low}$ & upper and lower energy buffer thresholds \\
\hline
\multicolumn{2}{|c|}{\bf Variables} \\
\hline
$\theta_{\rm COMM}(t)$ & total BSs energy consumption at time slot $t$\\
$\theta_{{\rm MEC}}(t)$ & server's energy consumption at time slot $t$\\
$\theta_{\rm CNT}(t)$ & energy drained due to the active containers,\\
    & w.r.t CPU utilization, at time slot $t$\\
$\theta_{\rm SC}(t)$ & energy drained due to container switching \\
   & the processing rates at time slot $t$\\
$\theta_{\rm OFF}(t)$ & energy induced by the TOE at $t$\\
$\theta_{\rm LNK}(t)$ & energy drained due to the \mbox{virtual-links}\\      
   & communication cost at time slot $t$\\
$\theta_{\rm DR}(t)$ & energy drained due to the number of active\\
  & transmission drivers at time slot $t$\\
$\theta_{\rm CC}(t)$ & total energy cost incurred by the content\\ 
     & caching process\\
$C(t)$ & number of containers to be active in time slot $t$\\
$f_{c}(t)$ &  instantaneous processing rate\\
$\psi_{c}(t)$ & load dependent factor\\
$r_c(t)$ & $c$-th virtual link communication rate at slot $t$\\
$\zeta_{n}(t)$ &  BS switching status indicator at $t$\\
$\theta_{\rm max}^{\rm NIC}(t)$ & maximum energy drained by the TOE at $t$\\
$\chi_c(t)$ & the expected processing time\\
$M(t)$ & number of active transmission drivers at $t$\\
$b(t)$ & energy buffer level in slot $t$\\
$H(t)$ & harvested energy profile in slot $t$ \\
$E(t)$ & purchased grid energy in slot $t$ \\
\hline
\end{tabulary}
\label{tab:variables}
\end{table}

\section{System Model}
\label{sec:sys}

In line with ETSI proposed MEC deployment scenarios discussed in~\cite{etsimec_access}, the considered network scenario is illustrated in Fig.~\ref{fig:eh_mec} above where the proposed model is \mbox{cache-enabled}, TCP/IP offload capable (i.e., enables {\it partial} offloading in the server's \ac{NIC} such as checksum computation~\cite{sohan2010characterizing}). The virtualized MEC server is assumed to be hosting $C$ containers deployed at an aggregation point, which is in proximity to a cluster of $N$ \acp{BS} from the same \ac{MN} operator. The \acp{BS} are interconnected to the \ac{MEC} server for computation workload offloading. Each network apparatus (BS, MEC server) is mainly powered by renewable energy harvested from wind and solar radiation, and it is equipped with an \ac{EB} for energy storage. In this case, energy can only be purchased from the grid supply to supplement the renewable energy supplies. The BSs coverage areas overlaps so that \mbox{green-based} load balancing is possible. The \ac{EM} is an entity responsible for selecting the appropriate energy source to fulfill the \ac{EB}, and also for monitoring the energy level of the \ac{EB}.  Then, the electromechanical switch (SW) aggregates the energy sources to fulfill the \ac{EB} level. In the MEC server, there is the presence of a virtualized access control router which acts as an access gateway for admission control, responsible for local and remote routing, and it is locally hosted as an application. Also, the MEC platform is able to track user location via the MEC Location Service Application Programmable Interface (LS API). Lastly, a \mbox{discrete-time} model is considered whereby time is discretized as $t = 1,2,\dots$ and each time slot $t$ has a fixed duration $\tau = \SI{30} {\minute}$. The list of symbols that are used in the paper is reported in Table~\ref{tab:variables}.

\subsection{Communication traffic and Energy consumption}
\label{sub:serverload}

\begin{figure}[t]
	\centering
	\resizebox{\columnwidth}{!}{\input{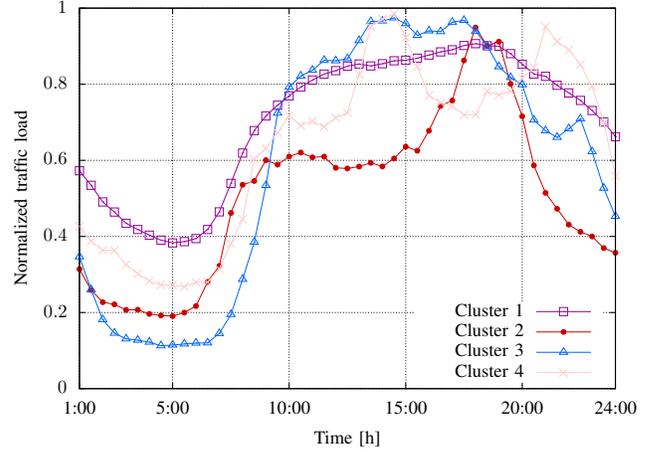}}
	\caption{Normalized BS traffic loads behavior represented as clusters. The data from~\cite{traces} has been split into four representative clusters.}
	\label{fig:trace_load}
\end{figure}

From a networking perspective, the understanding and characterization of the energy consumption within the \ac{MN} can pave the way towards more efficient and \mbox{user-oriented} networking solutions. This can be achieved through the use of historical mobile traffic traces such as Call Detail Records (CDRs) obtained from mobile operators, specifically in the \ac{EPC} network. 
Due to the difficulties in obtaining relevant open source datasets containing computing requests, real \ac{MN} traffic load traces obtained from the \ac{TIM} network (availed through the Big Data Challenge~\cite{bigdata2015tim}) are used to emulate the computational load.
In order to understand the daily traffic load patterns, the clustering algorithm \mbox{X-means}~\cite{pelleg2000x} has been applied to classify the load profiles into several categories. 
Here, each \ac{BS} $n$ is assumed to have a related load profile ${L}_{n}(t)$ which is picked at random as one of the four clusters in Fig.~\ref{fig:trace_load}. In addition, it is assumed that ${L}_{n}(t)$ consists of $80\%$ delay sensitive workloads $\gamma_{n}(t)$ and the remainder is delay tolerant workloads. The total aggregate delay sensitive workload per time instance is $\xi(t) = \sum_{n=1}^{N} \gamma_{n}(t)$.

The virtualized router in the MEC server of Fig.~\ref{fig:eh_mec} determines the amount of workload that can be accepted by the input buffer at slot $t$ and the aggregated (or admitted) workload is denoted by $L_{\rm in}(t) \in [0, L_{\rm in}^{\rm max}]$ (measured in [Mbits]). $L_{\rm in}^{\rm max}$ is the maximum input buffer size. 
In addition, it is assumed that the input/output (I/O) queue of the MEC server are \mbox{loss-free} and they implement the \mbox{First-In First-Out} (FIFO) service discipline, thus $L_{\rm in}(t) = L_{\rm out}(t)$, where $L_{\rm out}(t)$ is the amount of the aggregate computed workload at the output buffer. 

The total energy consumption ([$\SI{} {\joule}$]) for the edge system at time slot $t$ is formulated as follows, inspired by~\cite{comp_plus_comm_mec}\cite{vm_book}:
\begin{equation}
	\theta_{\rm EDGE}(t) = \theta_{\rm COMM}(t) +  \theta_{\rm MEC}(t) \, ,         
	\label{eq:mecconsupt}
\end{equation}
\noindent where $\theta_{\rm COMM}(t)$ is the energy consumption term induced by all BS communications and $\theta_{\rm MEC}(t)$ is the energy consumption term induced by the \ac{MEC} server's computing, caching and communication processes. \\

\subsubsection*{\bf BS energy cost} the overall energy consumption within the coverage area is defined as the sum of all the \acp{BS} components:
\begin{equation}
\label{eq:bs_cost}
\mbox{$\theta_{\rm COMM}(t) = \sum_{n = 1}^{N} \theta_{{\rm BS},n} (t) = \sum_{n = 1}^{N}(\delta_{n}(t) \theta_{0} + \theta_{{\rm load},n}(t))$},
\end{equation}
where $\delta_{n}(t) \in  \{0, 1\}$ is the BS $n$ switching status indicator ($1$ for {\it active mode} and $0$ for {\it power saving mode}), $\theta_{0}$ is a constant value (load independent) representing the operation energy which includes baseband processing, radio frequency power expenditures, etc. $\theta_{{\rm load},n}(t)$ is the load dependent \ac{BS} transmission power to the served users that guarantees low latency at the edge. It is obtained by using the transmission model in~\cite{mec_lyapunov}.\\

\subsubsection*{\bf MEC energy cost} the energy drained due to the computing, caching and communication processes is defined as:
\begin{equation}
    \begin{aligned}
    \theta_{\rm MEC}(t) & =  \theta_{\rm CNT}(t) + \theta_{\rm SWT}(t) + \theta_{\rm OFF}(t) \\
                           & + \theta_{\rm LNK}(t) + \theta_{\rm DR}(t) + \theta_{\rm CC}(t),
   \end{aligned}
   \label{eq:mec_cost}
\end{equation}
where $\theta_{\rm CNT}(t)$ is the energy drained due to the active containers, w.r.t \ac{CPU} utilization, and $\theta_{\rm SWT}(t)$ is the energy drained due to containers adapting to new processing rates $f_c(t) \in [f_0, f_{\rm max}]$ [(Mbit/s)].
The term $f_0$ is the zero speed of the container, e.g., deep sleep or shutdown, and $f_{\rm max}$ is the maximum available processing rate for container $c$. It is worth noting that actual containers are generally instantiated atop physical computing cores which offer only a finite set of processing speeds. 
The term $\theta_{\rm OFF}(t)$ is the energy induced by the TCP/IP offload on the NIC and $\theta_{\rm LNK}(t)$ is the energy drained due to the \mbox{virtual-links} (to-and-from containers) communication cost. Then, $\theta_{\rm DR}(t)$ is the amount of energy consumed by the active transmission drivers and $\theta_{\rm CC}(t)$ is the total energy cost incurred by the content caching process.

In this regard, it is assumed that \mbox{real-time} processing of computation workloads are performed in parallel over the containers interconnected by a  \mbox{rate-adaptive} Virtual LAN (VLAN). In addition, it is also assumed that the \ac{CPU} frequency is fixed at each user and may vary over users. 
The amount of energy consumed by the \ac{CPU} is related to the provisioned computing resources, i.e., the CPU share allocated to each container, per time instance $t$, named $C(t) \leq C$, index by $c$. Thus, $\theta_{\rm CNT}(t)$ is defined as~\cite{online_pimrc}: 
\begin{equation}
     \mbox{$\theta_{\rm CNT}(t) = \sum_{c=1}^{C(t)}\theta_{{\rm idle}, c}(t) + \psi_{c}(t) (\theta_{{\rm max},c}(t)-\theta_{{\rm idle}, c}(t))$},
    \label{eq:vm_cpu}
\end{equation}
where $\theta_{{\rm idle}, c}(t)$  represents the {\it static} energy drained by container $c$ in the idle state, \mbox{$\psi_{c}(t) = (f_{c}(t)/f_{\max})^{2}$} is the utilization function of container $c$~\cite{hayes_2004} and $\theta_{{\max}, c}(t)$ is the {\it maximum} energy that container $c$ can consume. The quantity $\psi_{c}(t) (\theta_{{\max}, c}(t) - \theta_{{\rm idle}, c}(t))$ represents the {\it dynamic} energy component of container $c$. 

The intelligent resource manager implements a suitable \mbox{frequency-scaling} policy in \mbox{real-time}, in order to allow the containers to scale up/down their processing rates $f_c(t)$ at the minimum cost. At this regard, it should be noted that switching from the processing frequency $f_c(t-1)$ (the processing rate at the $(t-1)$ time instance) to the next processing frequency $f_c(t)$ entails an energy cost, $\theta_{\rm SWT}(t)$. This depends on the absolute processing rate gap $|f_c(t) - f_c(t-1)|$, thus $\theta_{\rm SWT}(t)$ is defined as:
\begin{equation}
	\mbox{$\theta_{\rm SWT}(t) = \sum_{c=1}^{C(t)} z_e(f_c(t) - f_c(t-1))^2$},
	\label{eq:vm_sc}
\end{equation}
where $z_e$ is the the \mbox{per-container} reconfiguration cost caused by a \mbox{unit-size} frequency switching. Typically, $z_e$ is limited to a few hundreds of $\SI{} {\milli\joule}$ per $(\SI{} {\mega\hertz})^{2}$.

Before proceeding, it is worth noting the following: at the beginning of time slot $t$, the online algorithm adaptively allocates the available resources and then determine the containers that are demanded, $C(t)$, the size of the workload allocated to the container $c$, denoted by $\lambda_{c}(t)$, and $f_c(t)$ for container $c$ that will yield the desired or expected processing time, \mbox{$\chi_{c}(t) = \lambda_{c}(t)/f_{c}(t)$}. $\chi_c(t) \leq \Delta$, where $\Delta$ is the maximum \mbox{per-slot} and \mbox{per-container} processing time ([s]). Note that \mbox{$L_{\rm in}(t) = \sum_{c = 1}^{C(t)} \lambda_c(t)$} is the amount of computational workload admitted in the MEC server, by the router. The amount of the workload to be admitted \mbox{per-slot} shall be decided at the beginning of each time slot depending on the forecasted green energy, grid power to be purchased, and the expected computational workloads $\hat{L}_{\rm in}(t)$. Moreover, virtualization technologies specify the minimum and maximum amount of resources that can be allocated per container~\cite{migrationpower}, thus the maximum amount is denoted by $\lambda_{\rm max}$. Lastly, the container(s) provisioning and workload allocation is discussed in Section~\ref{sub:edge_manager}, {\it Remark 1}, and $f_c(t) \stackrel{\Delta}{=} \lambda_c(t) / \Delta$.

By implementing a TCP Offload Engine (TOE) in \mbox{high-speed} computing environments, some TCP/IP processing is offloaded to the network adapter for the purpose of reducing the CPU utilization.
To obtain the energy cost incurred, the performance measure for the Broadcom (Fibre) $10$ Gbps NIC~\cite{sohan2010characterizing} is considered here as an example of a TCP/IP \mbox{offload-capable} device. Note that $\theta_{\rm OFF}(t)$ is data volume dependent and it is obtained as:
\begin{equation}
    \mbox{$\theta_{\rm OFF}(t) = \zeta(t)\,\theta_{\rm idle}^{\rm NIC}(t) \, + \theta_{\rm max}^{\rm NIC}(t)$},
  \label{eq:vm_toa}
\end{equation} 
where $\theta_{\rm idle}^{\rm NIC}(t) > 0$ is the energy drained by the TOE when powered, with all links connected without any data transfer. This provides an opportunity for switching off the network adapter if there is no data transfer, making the energy drained to be zero. For this, $\zeta(t) = (0,1)$ is the switching status indicator ($1$ for active state and $0$ for idle state) and $\theta_{\rm max}^{\rm NIC}(t) = \frac{g(t)\cdot L_{\rm in}(t)}{\eta}$ is the maximum energy drained, where $g(t)$ is a fractional value representing the amount of load computed in the network adapter and $\eta$ is the NIC best throughput performance, hereby obtained as a fixed value measured in [Gbit/$\SI{} {\joule}$]. 

In order to keep the transmission delays from (to) the scheduler to (from) the connected containers at a minimum value, it is assumed that each container $c$ communicates with the resource scheduler through a dedicated reliable link that operates at the transmission rate of $r_c(t)$ [(bit/s)]. Thus, the energy needed for sustaining the \mbox{two-way} $c^{\rm th}$ link is defined as, inspired by~\cite{nicola}:
\begin{equation}
	\mbox{$\theta_{\rm LNK}(t) = 2\,\sum_{c=1}^{C(t)} P_c(r_c(t))(\lambda_c(t)/ r_c(t))$},
	\label{eq:vm_vlan}
\end{equation}
where \mbox{$P_c(r_c(t)) = S_c (2^{r_c(t)/W_c} - 1)$} is the power drained by the $c^{\rm th}$ communication link and \mbox{$S_c = \frac{W_c \times N_0^{(c)}}{g_c}$}. $N_0^{(c)}(\SI{} {\watt /\hertz})$ is the noise spectral power density, $W_c$ is the bandwidth, and $g_c$ is the (\mbox{non-negative}) gain of the $c^{\rm th}$ link. 
In practical application scenarios, the maximum \mbox{per-slot} communication rate within the \mbox{intra-VLAN} is generally limited up to an assigned value $r_{\rm max}$. Thus, the following hard constraint must hold: $ \sum_{c=1}^{C(t)} r_c(t) \leq r_{\rm max}$.

In this regard, a \mbox{two-way} per task execution delay is considered. Here, there is a total of $c = \{1,\dots, C(t)\}$ link connection delays, each denoted by $\varrho_c(t) = \lambda_c(t)/r_c(t)$, and $\chi_c(t) \leq \Delta$ where $\Delta$ is the server's response time, i.e., the maximum time allowed for processing the total computation load and it is fixed in advance regardless of the task size allocated to container $c$. Since parallel \mbox{real-time} processing is assumed in this work, the overall communication equates to $2\,\varrho_{c}(t) + \Delta$. Therefore, the hard \mbox{per-task} delay constraint on the computation time is: $\max \{2\,\varrho_{c}(t)\} + \Delta = \tau_{\rm max}$, where $\tau_{\rm max}$ is the maximum tolerable delay, which is fixed in advance.

Edge distributed devices utilize \mbox{low-level} signaling for information sharing. Thus, edge computing systems receives information from mobile devices within the local access network to discover their location. In return, for every {\it client} who offloaded their task into the \ac{MEC} server associated with the radio nodes, i.e., \acp{BS}, its location and the computation result is known through the \mbox{LS} (which is a service that supports UE’s location retrieval mechanism, and then passing the information to the authorized applications within the server), thus enabling the \mbox{location-aware} traffic routing and obtaining the number of transmission drivers to be active for data transfers.
The term $\theta_{\rm DR}(t)$ depends on the number of active laser (optical) drivers, named $M(t) \leq M$, where $M$ is the total number of drivers, that are required for transferring $ \ell_m(t) \in L_{\rm out}(t)$ in time slot $t$ ($\ell_m(t)$ is the downlink traffic volume ([bits] of the driver at slot $t$). $L_{\rm out}(t)$ is accumulated over a fixed period of time to form a {\it batch} at the output buffer. 
This means that a large number of drivers yield large transmission speed while at the same time resulting into high energy consumption~\cite{laser_tuning}. Therefore, the energy consumption can be minimized by launching an optimal number of drivers for the data transfer.

The energy drained during the data transmission process consists of the following: a constant energy for utilizing each fast tunable driver denoted by $d_m(t) ([\SI{} {\joule}$/s]), the target transmission rate $r_{0}$ [bits/s] and $ L_{\rm out}(t)$. Thus, the energy is, inspired by~\cite{comp_plus_comm_mec}:
\begin{equation}
	\mbox{$\theta_{\rm DR}(t) = \sum_{m=1}^{M(t)}\frac{d_m(t)\,l_m(t)}{r_0}$},
	\label{eq:vm_wcom}
\end{equation}
\noindent where the parameter $M(t)$ is obtained using the total number of target \acp{BS} as $M(t) = \ceil[\big]{\frac{1}{\textupsilon}\cdot(\frac{\omega (t) + 1}{\omega(t)})^2}$, where $\omega(t) = \sqrt{\frac{\rho}{\sigma N_{\rm BS}(t)}}$. $\textupsilon \in (0,1]$ is a controllable factor that determines the delay constraint of optical networks, $\sigma$ ([$\SI{} {\milli\second}$]) is the reconfiguration cost for tuning the transceivers, $N_{\rm BS}(t)$ is an integer value representing the total number of target \acp{BS} at time slot $t$, and $\rho$ is the number of time slots at which the computed workload is accumulated at the output buffer. Thus, the terms $\textupsilon, \sigma,$ and $\rho$ are fixed values, and $ L_{\rm out}(t)$ is equally distributed over the $M(t)$ drivers.

The \ac{MEC} server is able to cache contents from the internet and store the contents closer to mobile users. The caching process also contributes to the energy consumption in the server. The caching process is restricted to only viral content. For example, when a video becomes viral, users watching it share and talk about that video, which will be then requested by other users after a response time. Taking into account the internet users response time $\bar{\lambda} (t)$, this epidemic behavior can be modelled by the \mbox{self-excited} Hawkes condition Poisson process described in~\cite{large_youtube}: $\bar{\lambda} (t) = V(t) + \sum_{t_i \leq t} \Omega_i \, k_i(t-t_i)$, where $k_i(t)$ is the response time function, $\Omega_i$ is the number of potential viewers who will be influenced after $t_i$, which is the time when the user $i$ shared the video. The term $V(t)$ is added as a component to the model to capture the views that are not triggered by the epidemic effect.
The energy consumption is mainly contributed by content caching and data transmission processes, as such $\theta_{\rm CC}(t)$ is defined as:
\begin{equation}
 \theta_{\rm CC}(t) = \bar{\lambda} (t)\,(\theta_{\rm TR} (t) + \theta_{\rm CACHE}(t))\,,
 \label{eq:cache}
\end{equation}
where $\theta_{\rm TR} (t)$ is the power consumption due to transmission and $\theta_{\rm CACHE}(t)$ is the power consumption contributed by the caching process. 

\subsection{Energy Patterns and Storage}
\label{sub:eebuffer}

\begin{figure}[t]
	\centering
	\resizebox{\columnwidth}{!}{\input{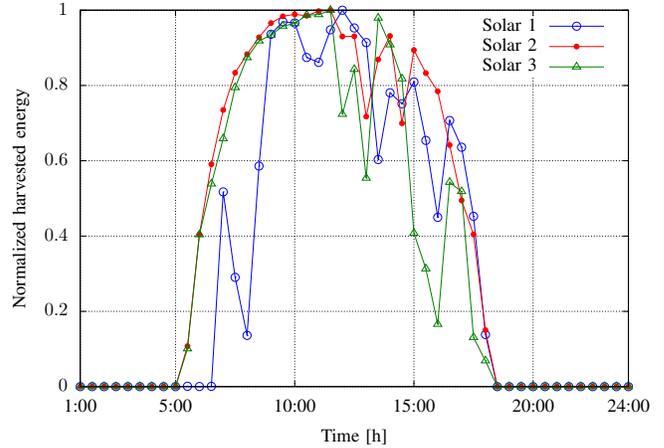}}
	\caption{Example traces for harvested solar traces and wind traces from~\cite{belgium}.}
	\label{fig:energy_trace}
\end{figure}

The rechargebale energy storage device is characterized by its finite energy storage capacity $b_{\rm max}$. At each time instance, the energy level reports are pushed from the \ac{BS} sites to the \ac{MEC} server. Thus, the \ac{EB} level $b(t)$ is known, enabling the provisioning of the required computation and communication resources, i.e., the required containers, transmission drivers and BSs to be active. In this paper, the amount of harvested energy $H(t)$, per site BS site, in time slot $t$ is obtained from \mbox{open-source} solar and wind traces from a farm located in Belgium~\cite{belgium} ({\it see} Fig.~\ref{fig:energy_trace} above). The data in the dataset matches our time slot duration ($\SI{30} {\minute}$). The dataset is the result of daily environmental records for a place assumed to be free from surrounding obstructions (e.g., buildings, shades). 

The harvested energy $H(t)$ is obtained by picking a day at random in the dataset and associating it with one site. Here, the wind energy is selected as a source during the solar energy \mbox{off-peak} periods. The available \ac{EB} level $b(t + 1)$ located at the BS site (BS $n$) or computing platform evolves according to the following dynamics: 
\begin{equation}
   \mbox{$b(t + 1) = \min\{b(t) + H(t) - \theta_{\rm site}(t) - a (t) + E(t), b_{\rm max}\}$},
\label{eq:offgrid}
\end{equation}
where $b (t)$ is the energy level in the battery at the beginning of time slot $t$, $\theta_{\rm site}(t)$ represent either $\theta_{{\rm BS},n}(t)$, the BS energy consumption of the communication site, or $\theta_{\rm MEC}(t)$, the energy drained at the computing platform, over time slot $t$, {\it see} Eq.~\eq{eq:bs_cost} and \eq{eq:mec_cost}. $a(t)$ is leakage energy and $E(t) \geq 0$ is the amount of energy purchased from the power grid.  Its worth noting that $b(t)$ is updated at the beginning of time slot $t$ whereas $H(t)$, $\theta_{{\rm BS},n}(t)$ and $\theta_{\rm MEC}(t)$, are  only known at the end of it. Thus, the energy constraint at the computing site must be satisfied for every time slot: $\theta_{\rm MEC}(t) \leq b(t)$.

For decision making in the GENM application, the received \ac{EB} level reports are compared with the following thresholds: $b_{\rm low}$ and $b_{\rm up}$, respectively termed the lower and the upper energy threshold with $0 < b_{\rm low} < b_{\rm up} < b_{\rm max}$. $b_{\rm up}$ corresponds to the desired energy buffer level at the BS site or computing site and $b_{\rm low}$ is the lowest EB level that any site should ever reach. If $b(t) < b_{\rm low}$, then BS $n$ or the computing site is said to be {\it energy deficient}. The suitable energy source at each time slot $t$ is selected based on the forecast expectations, i.e., the expected harvested energy $\hat{H}(t)$. If $\hat{H}(t)$ is enough to reach \mbox{$b_{\rm up}$}, no energy purchase is needed. Otherwise, the remaining amount up to \mbox{$b_{\rm up}$}, i.e., \mbox{$E(t) = b_{\rm up} - b(t)$}, is purchased from the electrical grid. Our optimization framework in Section~\ref{sub:opt_prob} makes sure that $b(t)$, never falls below $b_{\rm low}$ and guarantees that \mbox{$b_{\rm up}$} is reached at every time slot.

\section{Problem Formulation}
\label{sec:prob}

In this section, the optimization problem is formulated to obtain reduced energy consumption through \mbox{short-term} traffic load and harvested solar energy forecasting along with energy management procedures. The optimization problem is defined in Section~\ref{sub:opt_prob}, and the edge system management procedures are presented in Section~\ref{sub:edge_manager}.

\subsection{Optimization Problem}
\label{sub:opt_prob}

Our objective is to improve the overall energy savings of the edge system through BS power saving modes (i.e., \mbox{green-based} traffic load balancing), autoscaling of containers, contents caching and tuning of the transmission drivers, and also to guarantee the QoS within the network. Note that at the end of each time slot, the EB states are updated depending on the harvested energy and the consumed energy, thereby linking \mbox{per-time} slot problems across time.

To achieve our objective, two cost functions are defined, one captures the edge system energy consumption and the other, handles the \ac{QoS}. This is defined as: F1) $\theta_{\rm EDGE}(t)$, weighs the energy consumption due to transmission in the BSs and the \mbox{computing-plus-communication} activities in the MEC server. F2) a quadratic term $(\xi(t)-L_{\rm in}(t))^{2}$, which accounts for the \ac{QoS}. At this regard, it is worth noting that F1 tends to push the system towards \mbox{self-sustainability} solutions and F2 favors solutions where the delay sensitive load is entirely admitted in the MEC server by the router application, taking into account the expected energy to be harvested in the computing site. A weight $\Gamma = [0,1]$ is utilized to balance the two objectives F1 and F2. The corresponding (weighted) cost function is defined as:
\begin{equation}
\label{eq:Jfunc_2}
\begin{aligned}
J(\delta,\psi,M, t) & \stackrel{\Delta}{=} \overline{\Gamma}\theta_{\rm EDGE}(\delta_n(t),\{\psi_c(t)\}, M(t), t)\\
                   & + \Gamma(\xi(t) - L_{\rm in}(t))^2 \, ,
 \end{aligned}
\end{equation}
where $\overline{\Gamma } \stackrel{\Delta}{=} 1 - \Gamma$. 
Hence, starting from $t = 1$ (i.e., $t = 1,2, \dots, T$) as the current time slot and the finite horizon $T$, the following optimization problem is formulated as:
\begin{eqnarray}
        \label{eq:objt_2}
        \textbf{P1} & : & \min_{\mathcal{E}} \sum_{t=1}^T J(\delta,\psi,M, t)  \\
        && \hspace{-1.25cm}\mbox{subject to:} \nonumber \\
        {\rm A1} & : & \delta_{n}(t) \in \{\epsilon,1\}, \nonumber \\
        {\rm A2} & : & \beta \leq C(t) \leq C, \nonumber \\
        {\rm A3} & : & b(t) \geq b_{\rm low} , \nonumber \\ 
        {\rm A4} & : & 0 \leq f_{c}(t) \leq f_{\rm max}, \nonumber \\
        {\rm A5} & : & 0 \leq \lambda_{c}(t) \leq \lambda_{\rm max}, \nonumber \\
        {\rm A6} & : & \chi_{c}(t) \leq \Delta, \nonumber\\
        {\rm A7} & : & \mbox{$\sum_{c=1}^{C(t)} r_c(t) \leq r_{\rm max}$}, \nonumber\\
        {\rm A8} & : & \mbox{$\theta_{\rm MEC}(t) \leq b(t)$}, \nonumber\\
        {\rm A9} & : & \max \{2\,\varrho_{c}(t)\} + \Delta = \tau_{\rm max}, \quad t=1,\dots, T \, , \nonumber     
\end{eqnarray}
where the set of objective variables to be configured at slot $t$ in the BS system and MEC server is defined as \mbox{$\mathcal{E} \stackrel{\Delta}{=} \{\{\delta_{n}(t)\}, C(t), \{\psi_c(t)\}, \{P_c(t)\}, \{\lambda_c(t)\}, \zeta(t), M(t)\}$}. The setting  handles the transmission and computing-plus-communication activities. Constraint A1 specifies the BS operation status (either {\it power saving} or {\it active}), A2 forces the required number of containers, $C(t)$, to be always greater than or equal to a minimum number \mbox{$\beta \geq 1$}: the purpose of this is to be always able to handle mission critical communications. A3 makes sure that the \ac{EB} level is always above or equal to a preset threshold $\beta_{\rm low}$, to guarantee {\it energy \mbox{self-sustainability}} over time. Furthermore, A4 and A5, bound the maximum processing rate and workloads of each running container $c$, with $c = 1,\dots, C(t)$, respectively. Constraint A6 represents a \mbox{hard-limit} on the corresponding \mbox{per-slot} and \mbox{per-VM} processing time. A7 bounds the aggregate communication rate sustainable by the VLAN to $r_{\rm max}$ and A8 ensures that the  energy consumption at the computing site (due to the admitted computational workload) is bounded by the available energy in the EB. A9 forces the server to process the offloaded tasks within the set value $\tau_{\rm max}$. 

From the optimization problem \textbf{P1}, it could be noted that $J(\zeta,\psi,M, t)$ consists of a \mbox{non-convex} component defined in Eq.~\eq{eq:vm_vlan}, while the others are convex and \mbox{non-decreasing}. In this case, Eq.~\eq{eq:vm_vlan} can be converted into a convex function using \ac{GP} concept~\cite{geo_prog}, by introducing alternative variables and approximations. In this case, fixed parameters and approximations are introduced, i.e., $\mu_c, \nu_c$.
In the sequel, the index $t$ is dropped to improve readability. Thus, letting \mbox{$r_c = 2\,\lambda_c/(\tau_{\rm max}-\Delta)$} and then obtaining $P_c(r_c)$ in terms of $\lambda_c$ by rearranging the \mbox{Shannon-Hartley} expression and substituting the value of $r_c$: $\hat{P}_c (r_c)= \frac{((2\,\lambda_c/(\tau_{\rm max} - \Delta)) - \nu_c\,W_c)\ln 2}{\mu_c W_c} + \ln(N_0^{(c)})-\ln g_c$. From the \mbox{Shannon-Hartley} expression, the presence of the \mbox{\it log-sum-exp} function is observed as it has been proven to be convex in~\cite{gp_trick} and recall that \mbox{$P_c (r_c) = \exp (\hat{P}_c(r_c))$}.

To solve {\bf P1} in~\eq{eq:objt_2}, the \ac{LLC} principles~\cite{hayes_2004}\cite{chung1992limited}, \ac{GP} technique~\cite{geo_prog}, and heuristics, is used towards obtaining the feasible system control inputs $\varphi(t) = (\{\delta_{n}(t)\}, C(t), \{\psi_c(t)\}, \{P_c(t)\}, \{\lambda_c(t)\}, \zeta(t), M(t))$ for $t=1,\dots,T$. Note that~\eq{eq:objt_2} can iteratively be solved at any time slot $t \geq 1$, by just redefining the time horizon as $t^\prime = t, t+1, \dots, t+T-1$. 

\subsection{Edge System Management}
\label{sub:edge_manager}

In this subsection, a traffic load and energy harvesting prediction method and an online management algorithm are proposed to solve the previously stated problem {\rm P1}. 

\subsubsection{Traffic load and energy prediction}
\label{predict}

Given a time slot duration of $\tau =\SI{30} {\minute}$, the time series prediction is performed, i.e., the $T = 3$ estimates of $\hat{L}_{n}(t)$ and $\hat{H} (t)$ are obtained by using an \ac{LSTM} developed in Python using Keras deep learning libraries (Sequential, Dense) where the network has a \mbox{one-dimensional} ($1$D) subsequence of data, single feature, and multi-step for an output. The dataset is split as $70\%$ for training and $30\%$ for testing. The efficient Adam implementation of stochastic gradient descent and fit the model for $20$ epochs with a batch size of $4$ is used. As for the performance measure of the model, the \ac{RMSE} is used.

\subsubsection{Edge system dynamics}
\label{bs_server_dyn}

The system state vector at time $t$ is denoted by \mbox{$q(t) = (\delta(t),C(t),M(t),b(t))$}, which contains the number of active \acp{BS}, $\delta(t)$, number of active containers, $C(t)$, transmission drivers for data transfers, $M(t)$, and the EB level, $b(t)$. The input vector \mbox{$\varphi(t)=(\{\delta_{n}(t)\}, C(t), \{\psi_c(t)\}, \{P_c(t)\}, \{\lambda_c(t)\}, \zeta(t), M(t))$} drives the MEC server behavior (handles the joint switching on/off of BSs, autoscaling and reconfiguration of containers, and the tuning of transmission drivers) at time $t$. In this work, $\{P_c^{*}(t)\}$ is obtained with CVXOPT toolbox\footnote{M. Andersen and J. Dahl. CVXOPT: Python Software for Convex Programming, 2019. [Online]. Available: https://cvxopt.org/}, and $\{\lambda_c^{*}(t)\}$ is obtained by following the prodecure outlined in {\it remark} $1$. 

The system behavior is described by the \mbox{discrete-time} \mbox{state-space} equation, adopting the \ac{LLC} principles~\cite{llcprediction}\cite{hayes_2004}:
\begin{equation}
\ q(t + 1) = \phi(q(t), \varphi(t)) \, , 
\end{equation}
\noindent where  $\phi(\cdot)$ is a behavioral model that captures the relationship between $(q(t),\varphi(t))$, and the next state $q(t + 1)$. Note that this relationship accounts for the amount of energy drained $\theta_{\rm COMM}(t), \theta_{\rm MEC}(t)$, that harvested $H(t)$ and that purchased from the electrical grid $E(t)$, which together lead to the next buffer level $\beta(t+1)$ through Eq.~\eq{eq:offgrid}.
The \ac{GENM} algorithm, finds the best control action vector that yields the desired energy savings within the edge network. Specifically, for each time slot $t$, problem~\eq{eq:objt_2} is solved, obtaining control actions for the prediction horizon $T$. The control action that is applied at time $t$ is $\varphi^{*}(t)$, which is the first one in the retrieved control sequence. This control amounts to setting the number of active BSs, $\{\delta_{n}^{*}(t)\}$, number of instantiated containers, $C^*(t)$ (along with their obtained $\{\psi_c^{*}(t)\}$, $\{P_c^{*}(t)\}$, $\{\lambda_c^{*}(t)\}$ values), \ac{NIC} status to either active or not, $\zeta^{*}(t) \in (0,1)$, and the optimal transmission drivers, $M^{*}(t)$. The entire process is repeated every time slot $t$ when the controller can adjust the behavior given the new state information. 

State $q(t)$ and $\varphi(t)$ are respectively measured and applied at the beginning of the time slot $t$, whereas the offered load $L(t)$ and the harvested energy $H(t)$ are accumulated during the time slot and their value becomes known only by the end of it. This means that, being at the beginning of time slot $t$, the system state at the next time slot $t+1$ can only be estimated, which is formally written as:
\begin{equation}
       \hat{q}(t + 1) = \phi(q(t),\varphi(t)) \,.
       \label{eq:state_forecast}
\end{equation}


\noindent\textbf{Remark 1 (Container provisioning and load distribution)} for a fair provisioning of the computing resources, $C(t)$, and the expected workload allocation, $\hat{\xi}(t+1)$, a remark is presented. Firstly, each container can only compute an amount of up to $\lambda_{\max}$ and to meet the latency requirements, $C(t)$ is obtained as: \mbox{$C(t) = \ceil[\big] {(\hat{\xi}(t+1)/ \lambda_{\max})}$}, where $\ceil[\big] {\cdot}$ returns the nearest upper integer. Secondly, to distribute the workload among the $C(t)$ containers, a heuristic process splits the computational workload $\lambda_c(t) = \lambda_{\max}$ to the first $C(t)-1$ containers, and the remaining workload $\lambda_{c}(t) = \hat{\xi}(t+1) - (C(t)-1)\lambda_{\max}$ to the last one. 

\subsubsection{Edge system management framework} 
\label{framework}

in order to perform traffic load balancing using the green energy as a performance metric, a framework is defined that will identify the \ac{BS} to be dynamically switched off and then steer the traffic load towards those \acp{BS} with sufficient green energy. To do this, the available operating interval is defined as the ratio of the next time slot available green energy  and the expected total power consumption (recall that the BS load is forecasted), per BS site, as
\begin{equation}
   \ I_n(t) = \frac{b_n(t+1)}{\theta_{{\rm BS},n}(t+1)} \geq 1.
   \label{eq:op_interval}
\end{equation}
If $I_n(t) < 1$, the BS site will not have sufficient energy to handle the expected traffic and it becomes a potential BS to be switched off. In the case where $I_n(t) \geq 1$, the site energy will be sufficient to handle the expected traffic. The potential BS to be switched off, denoted by BS $n$, will offload its traffic load to a neighboring BS, denoted by BS $nn^\prime$. For BS $nn^\prime$ to be able to handle the offloaded traffic, the energy must be sufficient, thus the \mbox{green-based} operating interval is defined as
\begin{equation}
  x_{nn^\prime}(t) = \frac{b_{nn^\prime}(t+1)}{\theta_{{\rm BS},nn^\prime}(t+1)} \,,
  \label{eq:new_interval}
\end{equation}
where $\theta_{{\rm BS},nn^\prime}(t+1)$ is the total energy consumption of the BS site when the traffic load from the neighboring BS is combined with the expected load of the BS, and $b_{nn^\prime}(t+1)$ is the next time slot energy.
Next, the BS \mbox{wake-up} procedure is discussed.\\

\subsubsection*{BS wake-up procedure} to support BS \mbox{re-activation} commands, the UE location fingerprints that are obtained from the \ac{LS} \ac{API} are considered. 
The UE trajectory is assumed to be sequential, i.e., from BS-to-BS along same direction (while still associated with the MEC server) and this is represented as $i_{1}(t)\rightarrow i_{2}(t) \rightarrow \dots  \rightarrow i_{n}(t)$, where $ i_{n}(t)$ refers to the serving/target BS node $n$ in association with user $i$, at time slot $t$. When a BS node is switched off it goes into discontinuous reception cycle and configure a timer to awake and listen. Here, the MEC server, as BSs manager, send \mbox{wake-up} control information as \mbox{wake-up} signaling (the information is a single bit). The \mbox{wake-up} information is only sent during the listening period. A BS $n$ can be woken up only if it meets the following conditions: (i) $b_n(t+1) > b_{\rm low}$ and (ii) a group of UEs that are associated with the \ac{MEC} server are expected to receive the computed results via BS $n$ (their trajectory is towards BS $n$ as reported by the \ac{LS} in the MEC).\\

\subsubsection{Green-based Edge Network Management (GENM) Algorithm}
\label{alg}

in order for the algorithm to manage the BS system, deciding upon the allocation of their transmission resources, and also handling the computing and communication process, the best control action, \mbox{$\varphi(t)=(\{\delta_{n}(t)\}, C(t), \{\psi_c(t)\}, \{P_c(t)\}, \{\lambda_c(t)\}, \zeta(t), M(t))$}, that will yield the expected system behavior shall be obtained. 

\begin{small}
\begin{algorithm}[t]
\begin{tabular}{l l}
{\bf Input:}  & $q(t)$ (current state) \\
{\bf Output:} & $\varphi^{*}(t)$  (control input vector)\\
01:		& \hspace{-1cm} Parameter initialization\\
		& \hspace{-1cm} ${\mathcal S}(t) = \{q(t)\}$ \\
02:		& \hspace{-1cm} {\bf for} ($k$ within the prediction horizon of depth $T$) {\bf do}\\
		& \hspace{-1cm}\quad - $\hat{L}_{\rm in}(t+k)$:= forecast the workload  \\
		&\hspace{-1cm}\quad - $\hat{H}_n(t+k)$:= forecast the energy\\
		&\hspace{-1cm}\quad - $I_n(t+k)$:= operating interval of each BS\\
		&\hspace{-1cm}\quad - $x_{nn^\prime}(t+k)$:= green-based operating interval\\
		& \hspace{-1cm}\quad - ${\mathcal S}(t+k) = \emptyset$ \\
03:		& \hspace{-1cm}\quad {\bf for} (each $q(t)$ in ${\mathcal S}(t+k)$) {\bf do}\\
             & \hspace{-1cm}\qquad - generate all reachable states $\hat{q}(t+k)$\\
             & \hspace{-1cm}\qquad - ${\mathcal S}(t+k) = {\mathcal S}(t+k) \cup \{\hat{q}(t+k)\}$\\
04:		& \hspace{-1.1cm} \quad\quad {\bf for} (each $\hat{q}(t+k)$ in $\mathcal S(t+k)$) {\bf do}\\
            & \hspace{-1.1cm}\qquad\quad - calculate the corresponding $\theta_{\rm EDGE}(\hat{q}(t+k))$\\
            & \hspace{-1.1cm}\qquad\quad taking into account of $\kappa_n$ from $L_{\rm out}(t)$\\
		& \hspace{-1.1cm} \quad\quad {\bf end for}\\
		& \hspace{-1.1cm}\quad\quad {\bf end for}\\
		& \hspace{-1cm} \quad {\bf end for}\\
05:		& \hspace{-1cm} - obtain a sequence of reachable states yielding\\
        & \hspace{-1cm}\quad minimum energy cost\\	
06:		& \hspace{-1cm} {$\varphi^{*}(t):=$ control leading from $q(t)$ to $\hat{q}_{\min}$}\\
07:		& \hspace{-1cm} {\bf Return $\varphi^{*}(t)$}
\end{tabular}
\caption{GENM Algorithm Pseudocode}
\label{algo:genm}
\end{algorithm}
\end{small}

The edge network management algorithm pseudocode is outlined in Algorithm~\ref{algo:genm} above and it is based on the \ac{LLC} principles from~\cite{llcprediction}\cite{hayes_2004}.
Starting from the {\it initial state}, the controller constructs, in a \mbox{breadth-first} fashion, a tree comprising all possible future states up to the prediction depth $T$. 
The algorithm proceeds as follows: A search set $\mathcal S$ consisting of the current system state is initialized (line 01), and it is accumulated as the algorithm traverse through the tree (line 03), accounting for predictions, accumulated workloads at the output buffer, mobile devices trajectory $i_n(t)$, past outputs and controls, operating intervals. The set of states reached at every prediction depth $t+k$ is referred to as $\mathcal S(t+k)$ (line 02). 
Given $q(t)$, the workload $\hat{L}_{\rm in}(t+k)$ and harvested energy $\hat{H}(t+k)$ is estimated first, then obtain the operating intervals $I_n(t+k), x_{nn^\prime}(t+k)$ (line 02), and generate the next set of reachable control actions by applying the accepted workload $\xi(t+k)$, energy harvested and \mbox{green-based} operating interval (line 03).
The energy cost function corresponding to each generated state $\hat{q}(t+k)$ is then computed (line 04), where $\hat{q}(t+k)$ take into account of $\eta_n$ as observed from $L_{\rm out}(t)$. Once the prediction horizon is explored, a sequence of reachable states yielding minimum energy consumption is obtained (line 05). The control action $\varphi^{*}(t)$ corresponding to $\hat{q}(t+k)$ (the first state in this sequence) is provided as input to the system while the rest are discarded (line 06). The process is repeated at the beginning of each time slot $t$.

\subsubsection*{Algorithm Complexity}

The algorithm is executed at each time instance and the corresponding time complexity is obtained as follows. The time complexity associated with the computation of the $I_n(t)$ and $x_{nn^\prime}$ is linear with the size of the BS group $|N|$ interconnected to the MEC server. Next, the complexity associated with updating the load allocation for the active BSs is $|N| - 1$, which leads to $O(|N|^2)$. In the worst case scenario (no BS has been switched off), the total complexity is  $|N| q(t)\varphi(t)T$, which is linear in all variables, namely, number of BSs interconnected to the MEC server, number of system states, number of control actions, and time horizon $T$.

\section{Performance Evaluation}
\label{sec:results}

In this section, some selected numerical results for the scenario of Section~\ref{sec:sys} are shown. The parameters that were used in the simulations are listed in Table~\ref{tab_opt}.

\begin{figure}[t]
	\centering
	\resizebox{\columnwidth}{!}{\input{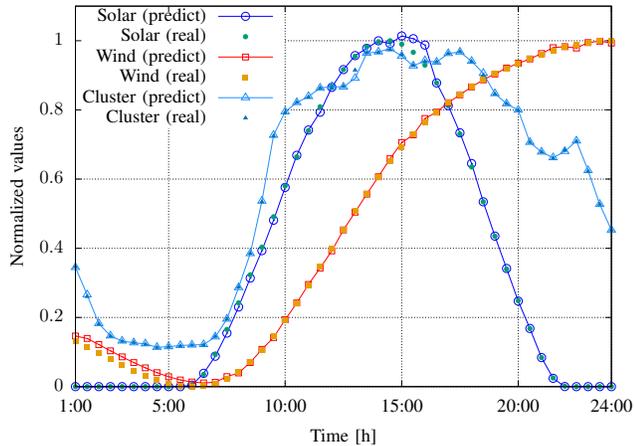}}
	\caption{One-step ahead predictive values for $L(t)$ and $H(t)$.}
	\label{fig:forecast}
\end{figure} 

\begin{table}
	\caption{System Parameters.}
	\center
	\begin{tabular} {|l| l|l|}
		\hline 
		{\bf Parameter} & {\bf Value} \\ 
		\hline
		Max. number of containers, $C$ &  $20$\\
		Min. number of containers, $\beta$ & $1$ \\
		Time slot duration, $\tau$ &  $\SI{30} {\minute}$\\
		Idle state energy for container $c, \theta_{{\rm idle}_c}(t)$ & $\SI{4} {\joule}$\\
		Max. energy for container $c, \theta_{{\rm max},m}(t)$ & $\SI{10} {\joule}$\\
		per-container reconfiguration cost, $z_e$ & $ 0.005 \rm J/(\rm MHz)^2$\\
		TOE in idle state, $\theta_{\rm idle}^{\rm NIC}(t)$ & $13.1 \rm J$\\
		Max. allowed processing time, $\Delta$ & $\SI{0.8} {\second}$\\
		Processing rate set, $\{f_c(t)\}$  & $\{0,50,70,90,105\}$\\
		Bandwidth, $W_c$ & $1 {\rm MHz}$\\
		Max. number of drivers, $M$ & $6$\\
		Noise spectral density, $N_0^{(c)}$ & $-174 \, {\rm dBm/Hz}$\\
		Max. container $c$ load, $\lambda_{\max}$ & $ 10$ MB\\
		NIC best performance throughput, $\eta$ & $1.4$ Gbit/J\\
		Driver energy, $d_m(t)$ & $1 \, \rm J/s$\\
		Target transmission rate, $r_0$ & $1 \, \rm Mbps$\\
		Controllable factor of delay, $\textupsilon$ & $0.96$\\
		Reconfiguration overhead, $\sigma$ & $20 \, \rm ms$\\
		Leakage energy, $a (t)$ & $2\, \mu \rm J$\\
		Energy storage capacity, $b_{\rm max}$ & $\SI{490} {\kilo\joule}$\\
		Lower energy threshold, $b_{\rm low}$  & $30$\% of $b_{\rm max}$\\
		Upper energy threshold, $b_{\rm up}$  & $70$\% of $b_{\rm max}$
\\
		\hline 
	\end{tabular}
	\label{tab_opt}
\end{table}

\subsection{Simulation Setup} 

A virtualized MEC server in proximity to a group of \acp{BS} is considered. The BS coverage areas overlaps to enable load balancing. Our time slot duration $\tau$ is set to $\SI{30} {\minute}$ and the time horizon is set to $T = 3$ time slots. For simulation, Python is used as the programming language.

\subsection{Numerical Results}

\textit{Data preparation:} The information from the used mobile and energy traces is aggregated to the set time slot duration. The mobile traces are aggregated from $\SI{10}{\minute}$ observation time to $\tau$. As for the wind and solar traces, they were aggregated from $\SI{15}{\minute}$ observation time to $\tau$. The used datasets are readily available in a public repository (\textit{see}~\cite{traces}).\\

In Fig.~\ref{fig:forecast}, the real and predicted values for BS traffic load and harvested energy is shown. Here, the forecasting routing tracks each value and predict it over \mbox{one-step}. The shown selected prediction results are for Cluster 3, Solar 3, and Wind 3. Then, Table~\ref{tab:pred} shows the the average RMSE of the normalized harvested energy and traffic load processes, for different time horizon values, $T \in \{1,2,3\}$. In the table, the term $H_{\rm wind} (t)$ represent the forecasted values for energy harvested from wind turbines and $H_{\rm solar} (t)$ is for the harvested energy from solar panels. From the obtained results, the prediction variations are observed between $H(t)$ and $L(t)$ when comparing the average RMSE. The measured accuracy is deemed good enough for the proposed optimization.

\begin{table}[t]
\footnotesize
\centering
\caption{Average prediction error (RMSE) for harvested energy and
traffic load processes, both normalized in [0,1].}
\begin{tabulary}{1.0\textwidth}{|L|L|L|L|}
\hline
  & {$T = 1$} & {$T = 2$} & {$T = 3$} \\
\hline
$L (t)$ & 0.010 & 0.013 & 0.018\\ \hline
$H_{\rm wind}(t)$   & 0.011 & 0.013 & 0.016\\ \hline
$H_{\rm solar}(t)$  & 0.010 & 0.011 & 0.014\\
\hline
\end{tabulary}
\label{tab:pred}
\end{table}

\begin{figure}[t]
	\centering
	\begin{subfigure}[t]{\columnwidth}
		\centering
		\resizebox{\columnwidth}{!}{\input{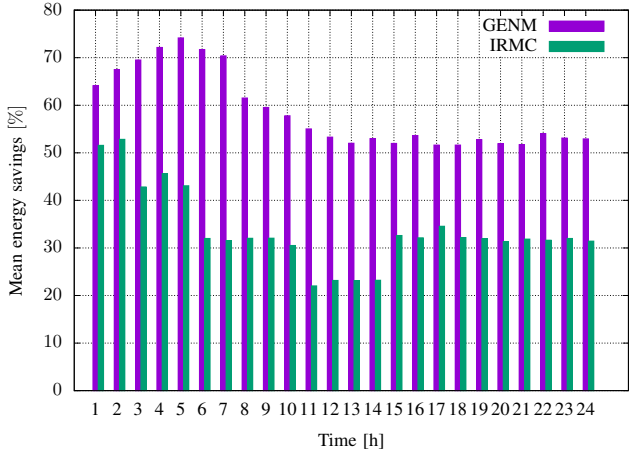}}
		\caption{Mean energy savings for $\Gamma = 0.5, |N| = 24, \lambda_{\rm max} = 10$ MB.}
		\label{fig:mec_en1}	
	\end{subfigure}
	\quad
	\begin{subfigure}[t]{\columnwidth}
		\centering
		\resizebox{\columnwidth}{!}{\input{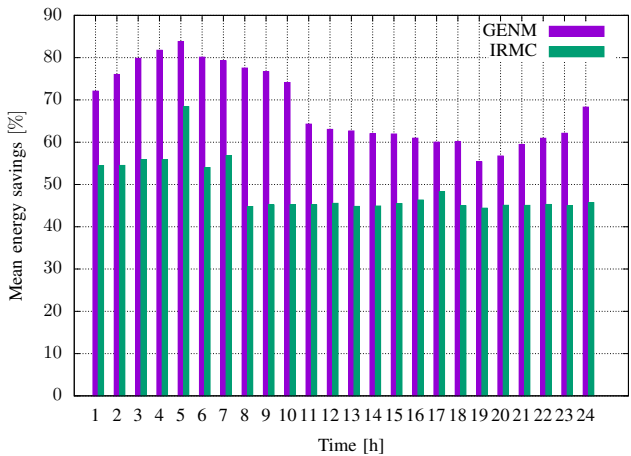}}
		\caption{Mean energy savings for $\Gamma = 0.5, |N| = 12, \lambda_{\rm max} = 10$ MB.}
		\label{fig:mec_en2}	
	\end{subfigure}
	\centering
	\caption{Mean energy savings within the MEC server}
	\label{figure:mecsavings}
\end{figure}

The GENM algorithm is benchmarked with another one, named \mbox{Iterative-based} Resource Manager with network impact Capability (IRMC), which is inspired by the iterative approach for computing platforms from~\cite{shojafar2015energy} and the use of the network impact towards load balancing from~\cite{E_oh}. Both algorithms make use of the learned information. Figs.~\ref{fig:mec_en1} and~\ref{fig:mec_en2} show the average energy savings obtained by GENM in the MEC server. In Fig.~\ref{fig:mec_en1}, the average results for GENM ($z_e = 0.005, |N| = 24, \Gamma = 0.5, \lambda_{\rm max} = 10$ MB) show energy savings of $59 \%$, while IRMC achieves $34 \%$ on average. As expected, the highest energy savings gain is observed in the early hours of the day ($\SI{1}{\hour} - \SI{8}{\hour}$) as the aggregated computational workload was at its lowest.
In Fig.~\ref{fig:mec_en2}, the average energy savings obtained by GENM is $68\%$ ($z_e = 0.005, |N| = 12, \Gamma = 0.5, \lambda_{\rm max} = 10$ MB) and for IRMC is $49 \%$. Again, here the highest peaks for energy savings are obtained from $\SI{1}{\hour} - \SI{8}{\hour}$.
The results are obtained with respect to the case where no energy management procedures are applied; i.e., the MEC server provisions the computing resources for maximum expected computation workload (maximum value of $\theta_{\rm MEC} (t), C = 20, \forall t$).
Comparing the results of Fig.~\ref{fig:mec_en1} and~\ref{fig:mec_en2}, we observed that when the BSs being manage by the MEC server are reduced (i.e., $12 < 24$), the aggregated delay sensitive workload is also reduced and this translates to reduced computation process demands, which in turn results into high energy savings.

\begin{figure}[t]
	\centering
	\resizebox{\columnwidth}{!}{\input{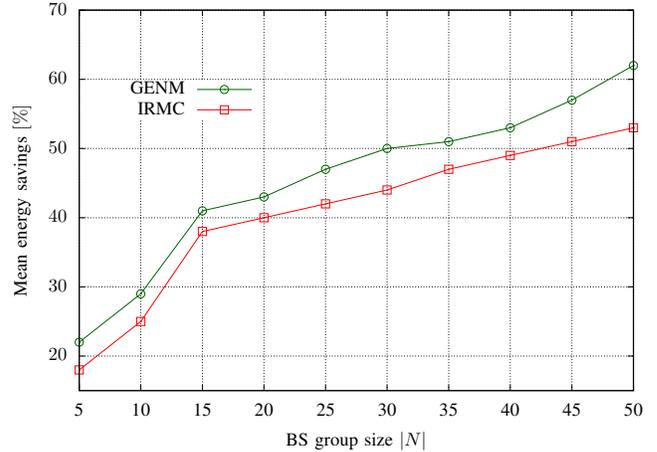}}
	\caption{Energy savings versus BS group size.}
	\label{fig:bsgroup}
\end{figure} 

Fig.~\ref{fig:bsgroup} shows the average energy savings obtained when green energy is used as a performance metric towards load balancing within a group of BSs. Here, the group size is increased from $|N| = 5$ to $50$, using incremental step size of $5$. The obtained energy savings are with respect to the case where all BSs are dimensioned for maximum expected capacity (maximum value of $\theta_{\rm COMM} (t)$). From the results, it is observed that the energy savings increase as the BS cluster grows, thanks to the load balancing among active BSs.

\begin{figure}[t]
	\centering
	\resizebox{\columnwidth}{!}{\input{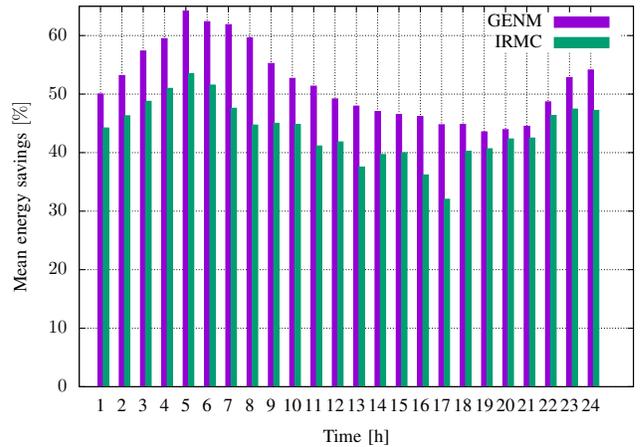}}
	\caption{Mean energy savings for the edge system.}
	\label{fig:edge_sys}
\end{figure} 

Then, Fig.~\ref{fig:edge_sys} shows the average energy savings for the edge system. Here, the BS group size is set to $|N| = 40$ and the obtained energy savings results are with respect to the case where no energy management procedures are applied, i.e., the BSs are dimensioned for maximum expected capacity (maximum value of $\theta_{\rm COMM} (t)$, $\forall t$) and the MEC server provisions the computing resources for maximum expected computation workload (maximum value of $\theta_{\rm MEC} (t)$, with $C = 20\,  \text{containers}, \forall t$). The average results of GENM ($z_e = 0.05, \lambda_{\rm max} = 10$ MB, $\Gamma = 0.5$) show energy savings of $51 \%$, while IRMC  achieves $44 \%$ on average. The effectiveness of the joint dynamic BSs management, autoscaling and reconfiguration of the computing resources, and on/off switching of the fast tunable laser drivers, coupled with foresighted optimization is observed in the obtained numerical results.

\section{Conclusions}
\label{sec:concl}

This paper envisioned an edge system where a group of \acp{BS} is placed in proximity to a \ac{MEC} server for ease of handling the offloaded computational workload and \acp{BS} management, and also the edge apparatuses are power by hybrid supplies, i.e., green energy is used in order to promote energy \mbox{self-sustainability} within the network and as a performance metric for traffic load balancing. The extra energy can only be purchased from the grid supply to supplement the renewable energy supplies. The considered energy cost model takes into account the computing, caching and communication processes within the MEC server, and \mbox{transmission-related} energy consumption in BSs. To intelligently manage the edge system, an online algorithm based on forecasting, control theory and heuristics, is proposed with the goal of minimizing the overall energy consumption and guarantee the quality of service within the network. 
The algorithm jointly performs (i) dynamic BS management using green energy as a performance metric, (ii) autoscaling and reconfiguration of the computing resources, workload and processing rate allocation, and lastly, (iii) switching on/off of fast tunable drivers. Numerical results, obtained with \mbox{real-world} energy and traffic load traces, demonstrate that the proposed algorithm achieves energy savings of above $50\%$ with respect to the allocated maximum \mbox{per-container} loads of $10$ MB. 
The computing platform is able to achieve energy savings from $59 \%$ to $68 \%$, depending on the size of the BS cluster.
The energy saving results are obtained with respect to the case where no energy management techniques are applied in the BS system and the MEC server.

\section*{Data Availability}

In this paper, open source datasets for the mobile network (MN) traffic load, solar and wind energy have been used. The details are as follows: (1) the real MN traffic load traces used to support the findings of this study were obtained  from the Big Data Challenge organized by Telecom Italia Mobile (TIM) and the data repository has been cited in this article. (2) The real solar and wind traces used to support the findings of this study have also been cited in this article.

\bibliographystyle{IEEEtran}
\scriptsize
\bibliography{biblio_new}

\begin{thebibliography}{10}
\providecommand{\url}[1]{#1}
\csname url@samestyle\endcsname
\providecommand{\newblock}{\relax}
\providecommand{\bibinfo}[2]{#2}
\providecommand{\BIBentrySTDinterwordspacing}{\spaceskip=0pt\relax}
\providecommand{\BIBentryALTinterwordstretchfactor}{4}
\providecommand{\BIBentryALTinterwordspacing}{\spaceskip=\fontdimen2\font plus
\BIBentryALTinterwordstretchfactor\fontdimen3\font minus
  \fontdimen4\font\relax}
\providecommand{\BIBforeignlanguage}[2]{{%
\expandafter\ifx\csname l@#1\endcsname\relax
\typeout{** WARNING: IEEEtran.bst: No hyphenation pattern has been}%
\typeout{** loaded for the language `#1'. Using the pattern for}%
\typeout{** the default language instead.}%
\else
\language=\csname l@#1\endcsname
\fi
#2}}
\providecommand{\BIBdecl}{\relax}
\BIBdecl

\bibitem{morabito_container}
R.~Morabito, V.~Cozzolino, A.~Y. Ding, N.~Beijar, and J.~Ott, ``{Consolidate
  IoT Edge Computing with Lightweight Virtualization},'' \emph{IEEE Network},
  vol.~32, no.~1, pp. 102--111, 2018.

\bibitem{interdigital}
``{Software-Defined and Cloud-Native Foundations for 5G Networks},''
  InterDigital, Denver, USA, Tech. Rep., May 2019.

\bibitem{green_balancing}
T.~Han and N.~Ansari, ``{A traffic load balancing framework for
  software-defined radio access networks powered by hybrid energy sources},''
  \emph{IEEE/ACM Transactions on Networking}, vol.~24, no.~2, pp. 1038--1051,
  2016.

\bibitem{globe_balancing}
J.~Xu, H.~Wu, L.~Chen, C.~Shen, and W.~Wen, ``{Online Geographical Load
  Balancing for Mobile Edge Computing with Energy Harvesting},'' \emph{arXiv
  preprint arXiv:1704.00107}, 2017.

\bibitem{E_oh}
E.~Oh, K.~Son, and B.~Krishnamachari, ``{Dynamic Base Station Switching-On/Off
  Strategies for Green Cellular Networks},'' \emph{IEEE Transactions on
  Wireless Communications}, vol.~12, no.~5, pp. 2126--2136, 2013.

\bibitem{edge_controller}
T.~Dlamini, {\'A}.~F. Gamb\'{\i}n, D.~Munaretto, and M.~Rossi, ``{Online
  Supervisory Control and Resource Management for Energy Harvesting BS Sites
  Empowered with Computation Capabilities},'' \emph{Wireless Communications and
  Mobile Computing}, 2019.

\bibitem{virttech}
R.~Morabito, ``{Power Consumption of Virtualization Technologies: An Empirical
  Investigation},'' in \emph{{IEEE International Conference on Utility and
  Cloud Computing (UCC)}}, Limassol, Cyprus, Dec 2015.

\bibitem{eempirical}
Y.~Jin, Y.~Wen, and Q.~Chen, ``{Energy efficiency and server virtualization in
  data centers: An empirical investigation},'' in \emph{{IEEE Conference on
  Computer Communications Workshops (INFOCOM Workshops)}}, Orlando, USA, Mar
  2012.

\bibitem{tbook}
\BIBentryALTinterwordspacing
T.~Dlamini, \emph{{Softwarization in Future Mobile Networks and Energy
  Efficient Networks}}, Nov. 2019. [Online]. Available:
  \url{https://www.intechopen.com/online-first/softwarization-in-future-mobile-networks-and-energy-efficient-networks}
\BIBentrySTDinterwordspacing

\bibitem{link_drivers}
S.~Fu, H.~Wen, J.~Wu, and B.~Wu, ``{Cross-Networks Energy Efficiency Tradeoff:
  From Wired Networks to Wireless Networks},'' \emph{IEEE Access}, vol.~5, pp.
  15--26, 2017.

\bibitem{comp_plus_comm_mec}
T.~Dlamini and A.~F. Gambin, ``{Adaptive Resource Management for a Virtualized
  Computing Platform in Edge Computing},'' in \emph{{IEEE International
  Conference on Sensing, Communication and Networking (SECON)}}, {Boston, USA},
  June 2019.

\bibitem{vm_book}
M.~Portnoy, \emph{{Virtualization essentials}}.\hskip 1em plus 0.5em minus
  0.4em\relax John Wiley and Sons, 2012.

\bibitem{oh2011toward}
E.~Oh, B.~Krishnamachari, X.~Liu, and Z.~Niu, ``{Toward dynamic
  energy-efficient operation of cellular network infrastructure},'' \emph{IEEE
  Communications Magazine}, vol.~49, no.~6, 2011.

\bibitem{llcprediction}
S.~Abdelwahed, N.~Kandasamy, and S.~Neema, ``{Online control for
  self-management in computing systems},'' in \emph{{IEEE Real-Time and
  Embedded Technology and Applications Symposium (RTAS)}}, {Ontario, Canada},
  May 2004.

\bibitem{bousia2016multiobjective}
A.~Bousia, E.~Kartsakli, A.~Antonopoulos, L.~Alonso, and C.~Verikoukis,
  ``{Multiobjective auction-based switching-off scheme in heterogeneous
  networks: To bid or not to bid?}'' \emph{IEEE Transactions on Vehicular
  Technology}, vol.~65, no.~11, pp. 9168--9180, 2016.

\bibitem{Han2013}
T.~Han and N.~Ansari, ``{On Optimizing Green Energy Utilization for Cellular
  Networks with Hybrid Energy Supplies},'' \emph{IEEE Transactions on Wireless
  Communications}, vol.~12, no.~8, pp. 3872--3882, 2013.

\bibitem{chen2018computation}
{Chen, Lixing and Zhou, Sheng and Xu, Jie}, ``{Computation peer offloading for
  energy-constrained mobile edge computing in small-cell networks},''
  \emph{{IEEE/ACM Transactions on Networking}}, vol.~26, no.~4, pp. 1619--1632,
  2018.

\bibitem{xu2016online}
X.~Jie and R.~Shaolei, ``{Online Learning for Offloading and Autoscaling in
  Renewable-Powered Mobile Edge Computing},'' in \emph{{IEEE Global
  Communications Conference (GLOBECOM) }}, Washington, USA, Dec. 2012.

\bibitem{online_pimrc}
T.~Dlamini, {\'A}.~F. Gamb\'{\i}n, D.~Munaretto, and M.~Rossi, ``{Online
  Resource Management in Energy Harvesting BS Sites through Prediction and
  Soft-Scaling of Computing Resources},'' in \emph{{IEEE PIMRC }}, {Bologna,
  Italy}, Sep 2018.

\bibitem{shojafar2015energy}
M.~Shojafar, N.~Cordeschi, D.~Amendola, and E.~Baccarelli, ``{Energy-saving
  adaptive computing and traffic engineering for real-time-service data
  centers},'' in \emph{{IEEE International Conference on Communication Workshop
  (ICCW)}}, London, UK, Jun 2015.

\bibitem{vm_char}
M.~Shojafar, N.~Cordeschi, and E.~Baccarelli, ``{Energy-efficient Adaptive
  Resource Management for Real-time Vehicular Cloud Services},'' \emph{IEEE
  Transactions on Cloud Computing}, 2016.

\bibitem{delay}
M.~Mukherjee, V.~Kumar, S.~Kumar, R.~Matam, C.~X. Mavromoustakis, Q.~Zhang,
  M.~Shojafar, and G.~Mastorakis, ``{Computation Offloading Strategy in
  Heterogeneous Fog Computing with Energy and Delay Constraints},'' in
  \emph{{IEEE International Conference on Communications (ICC)}}, {Dublin,
  Ireland}, June 2020.

\bibitem{jamil2020job}
{Jamil Bushra and Shojafar Mohammad and Ahmed Israr and Ullah Atta and Munir
  Kashif and Ijaz Humaira}, ``{A job scheduling algorithm for delay and
  performance optimization in fog computing},'' \emph{{Concurrency and
  Computation: Practice and Experience}}, vol.~32, no.~7, 2020.

\bibitem{mukherjee2019joint}
M.~Mithun, K.~Suman, S.~Mohammad, Z.~Qi, and X.~{Mavromoustakis Constandinos},
  ``{Joint task offloading and resource allocation for delay-sensitive fog
  networks},'' in \emph{{IEEE International Conference on Communications
  (ICC)}}, {Shanghai, China}, May 2019.

\bibitem{task_coord}
T.~Zhao, S.~Zhou, X.~Guo, and Z.~Niu, ``{Tasks scheduling and resource
  allocation in heterogeneous cloud for delay-bounded mobile edge computing},''
  in \emph{{IEEE International Conference on Communications (ICC)}}, {Paris,
  France}, May 2017.

\bibitem{laser_tuning}
B.~Wu, S.~Fu, X.~Jiang, and H.~Wen, ``{Joint Scheduling and Routing for QoS
  Guaranteed Packet Transmission in Energy Efficient Reconfigurable WDM Mesh
  Networks},'' \emph{IEEE Journal on Selected Areas in Communications},
  vol.~32, no.~8, pp. 1533--1541, 2014.

\bibitem{llc_datacenter}
D.~Kusic, J.~O. Kephart, J.~E. Hanson, N.~Kandasamy, and G.~Jiang, ``{Power and
  Performance Management of Virtualized Computing Environments Via Lookahead
  Control},'' in \emph{{International Conference on Autonomic Computing}},
  {Chicago, USA}, Jun. 2008.

\bibitem{hayes_2004}
J.~P. Hayes, ``{Self-Optimization in Computer Systems via On-Line Control:
  Application to Power Management},'' in \emph{{Proceedings of the First
  International Conference on Autonomic Computing}}, {Washington,USA}, May
  2004.

\bibitem{forecasting}
R.~Hyndman and G.~Athanasopoulos, \emph{{Forecasting: principles and
  practice}}.\hskip 1em plus 0.5em minus 0.4em\relax OTexts: Melbourne,
  Australia, 2013.

\bibitem{lstmlearn}
I.~Goodfellow, Y.~Bengio, and A.~Courville, \emph{{Deep Learning}}.\hskip 1em
  plus 0.5em minus 0.4em\relax MIT Press, 2016.

\bibitem{chung1992limited}
S.-L. Chung, S.~Lafortune, and F.~Lin, ``{Limited lookahead policies in
  supervisory control of discrete event systems},'' \emph{IEEE Transactions on
  Automatic Control}, vol.~37, pp. 1921--1935, 1992.

\bibitem{ferdowsi2017deep}
\BIBentryALTinterwordspacing
A.~Ferdowsi, U.~Challita, and W.~Saad, ``{Deep Learning for Reliable Mobile
  Edge Analytics in Intelligent Transportation Systems},'' 2017. [Online].
  Available: \url{arXiv preprint arXiv:1712.04135}
\BIBentrySTDinterwordspacing

\bibitem{kumar2018long}
J.~Kumar, R.~Goomer, and A.~K. Singh, ``{Long Short Term Memory Recurrent
  Neural Network (LSTM-RNN) Based Workload Forecasting Model For Cloud
  Datacenters},'' \emph{Procedia Computer Science}, vol. 125, pp. 676--682,
  2018.

\bibitem{etsimec_access}
S.~Kekki, W.~Featherstone, Y.~Fang, P.~Kuure, A.~Li, A.~Ranjan, D.~Purkayastha,
  F.~Jiangping, D.~Frydman, G.~Verin, K.~Wen, K.~Kim, R.~Arora, A.~Odgers,
  L.~M. Contreras, and S.~Scarpina, ``{MEC in 5G Networks},'' ETSI,
  Sophia-Antipolis, France, Tech. Rep., Jun 2018.

\bibitem{sohan2010characterizing}
S.~Ripduman, R.~Andrew, A.~W. Moore, and M.~Kieran, ``{Characterizing 10 Gbps
  network interface energy consumption},'' in \emph{{IEEE 35th Conference on
  Local Computer Networks (LCN)}}, Colorado, USA, Oct 2010.

\bibitem{bigdata2015tim}
\BIBentryALTinterwordspacing
{Open Big Data Challenge}. [Online]. Available:
  \url{https://dandelion.eu/datamine/open-big-data/}
\BIBentrySTDinterwordspacing

\bibitem{pelleg2000x}
D.~Pelleg, A.~W. Moore \emph{et~al.}, ``{X-means: Extending K-means with
  efficient estimation of the number of clusters},'' in \emph{{Proceedings of
  the Seventeenth International Conference on Machine Learning (ICML)}}, {San
  Francisco, USA}, Jun 2000.

\bibitem{mec_lyapunov}
L.~Chen, S.~Zhou, and J.~Xu, ``{Energy Efficient Mobile Edge Computing in Dense
  Cellular Networks},'' in \emph{{IEEE International Conference on
  Communications (ICC)}}, Paris, France, May 2017.

\bibitem{traces}
\BIBentryALTinterwordspacing
{Mobile and Energy datasets}. [Online]. Available:
  \url{https://github.com/lihles/mobile-datasets}
\BIBentrySTDinterwordspacing

\bibitem{migrationpower}
M.~Cardosa, M.~R. Korupolu, and A.~Singh, ``{Shares and utilities based power
  consolidation in virtualized server environments},'' in \emph{{IFIP/IEEE
  International Symposium on Integrated Network Management}}, New York, USA,
  June 2009.

\bibitem{nicola}
N.~Cordeshi, M.~Shojafar, and E.~Baccarelli, ``{Energy-saving self-configuring
  network data centers},'' \emph{{Computer Networks}}, vol.~57, no.~17, pp.
  3479--3491, 2013.

\bibitem{large_youtube}
F.~B. Abdesslem and A.~Lindgren, ``{Large scale characterisation of YouTube
  requests in a cellular network},'' in \emph{{Proceeding of IEEE International
  Symposium on a World of Wireless, Mobile and Multimedia Networks}}, Sydney,
  Australia, Jun 2014.

\bibitem{belgium}
\BIBentryALTinterwordspacing
``{Generation Data (Solar and Wind)}.'' [Online]. Available:
  \url{https://www.elia.be/en/grid-data/power-generation}
\BIBentrySTDinterwordspacing

\bibitem{geo_prog}
W.-C. Ho, L.-P. Tung, T.-S. Chang, and K.-T. Feng, ``{Enhanced component
  carrier selection and power allocation in LTE-advanced downlink systems},''
  in \emph{{2013 IEEE Wireless Communications and Networking Conference
  (WCNC)}}, Shanghai, China, April 2013.

\bibitem{gp_trick}
S.~Boyd and L.~Vandenberghe, \emph{{Convex Optimization}}.\hskip 1em plus 0.5em
  minus 0.4em\relax Cambridge University Press, 2004.

\end{thebibliography}
\end{document}